\documentclass[11pt,letterpaper,fleqn]{article}

\usepackage{graphicx}
\usepackage{tabulary}
\usepackage{lineno}
\usepackage{hyperref}
\usepackage{subfigure}
\usepackage{cite}
\usepackage{xcolor,colortbl}
\usepackage{wrapfig}
\usepackage{enumitem} 
\usepackage{titlesec}
\usepackage{longtable}
\title{Kubernetes Deployment Options for On-Prem Clusters}
\author{Lincoln Bryant, Robert W. Gardner, Fengping Hu, David Jordan  \\ \emph{\small Enrico Fermi Institute,} \emph{\small University of Chicago} \\ Ryan P. Taylor \\ \emph{\small University of Victoria} } 
\date{June 2024}

\begin{document}

\maketitle

\section{Introduction}
First released by Google over a decade ago, Kubernetes\cite{Kubernetes2} is an open-source service orchestration platform built to address the challenges of running containerized workloads at scale. Kubernetes is an open successor to an internal container orchestration system, Borg\cite{borg}, that manages the application lifecycle of many of Google's most popular products today. This shift to an open ecosystem is facilitated by a vendor-neutral oversight and support organization, the Cloud Native Computing Foundation (CNCF)~\cite{cncf}, a part of the Linux Foundation. The CNCF hosts core pieces of the Kubernetes development infrastructure, promulgates the cloud-native computing definition and technical vision through a Technical Oversight Comittee~\cite{cncf_toc}, and organizes industry-wide conferences (such as KubeCon and Cloud Native Con).  This has led to a thriving landscape of cloud native software products covering application definition and development, orchestration and management, runtime and provisioning products~\cite{cncf_landscape}. 

Kubernetes is now used, maintained, and extended by hundreds of companies world-wide with thousands of contributors and end-users. It also is growing in popularity in the field of high energy physics computing, with an increasing number of infrastructures managed by the platform~\cite{second_k8s_hep_meetup, ATLAS:2020frn, Taylor:2024cja, megino2024operational, cern_kubernetes, Albin:2023szc, TEDESCHI2024108965}. An un-opinionated and flexible architecture has allowed Kubernetes to become a ubiquitous, vendor-neutral platform for running a variety of workloads. In cloud computing, this has allowed vendors to tailor their Kubernetes offerings in a way that best fits hyperscale infrastructure while abstracting away the details of vendor-specific storage, networking and instancing from users. This is also a significant advantage for developers because the de-facto standardized abstractions provided by Kubernetes help avoid vendor lock-in and contribute significantly to a flourishing ecosystem.  

However, system administrators who plan to deploy their own cluster on bare metal or otherwise on-premise may find the flexible and pluggable design of Kubernetes to be challenging. To deploy a functioning cluster into production, administrators must choose appropriate tools to provide essential networking, storage, security and monitoring among other items. To ease the difficulty of constructing a production Kubernetes cluster, a number of opinionated approaches have coalesced in the broader community. Indeed, Kubernetes is often likened to a ``kernel of a distributed operating system''. To extend this metaphor, efforts to provide fully integrated Kubernetes deployments should perhaps be seen as Kubernetes distributions, much in the same way as Debian or Red Hat are distributions in the Linux environment.

\section{Deployment of Kubernetes Distributions}

The following sections describe three different Kubernetes distributions, including their deployment method and any additional steps needed to bring a cluster into production. This note covers kubeadm~\cite{kubeadm} and Kubespray~\cite{kubespray}, Red Hat's OKD~\cite{okd} and its commercially supported downstream version OpenShift, as well as SUSE's Rancher~\cite{rancher} which allows for deployment via the light-weight K3S~\cite{k3s} distribution as well as RKE2~\cite{rancher_gov_rke2}, a distribution focused on the US Federal Government sector.

\subsection{\emph{kubeadm}}
Following the metaphor of Linux distributions, the
\emph{kubeadm} tool shipped by the Kubernetes authors is most similar to a distribution that ships only a minimal base system with no ``batteries included'' to allow for maximum flexibility. While not completely manual, \emph{kubeadm} is designed to quickly instantiate the minimum viable Kubernetes cluster. The authors intend\cite{kubernetes-kubeadm-create-cluster}  for \emph{kubeadm} to be an un-opinionated building block used by higher-level tools to construct clusters.  One such high-level tool, Kubespray, will be examined later in this article.

\subsubsection{Control plane initialization}

When creating a Kubernetes cluster via \emph{kubeadm}, the Kubernetes control plane is initialized in the following way:
\begin{itemize}
    \item The \emph{kubelet} daemon is started on the control plane host(s).
    \item Certificates are issued for intra-cluster communication.
    \item The distributed key-value store \emph{etcd} is instantiated to hold cluster state.
    \item The Kubernetes API service is started as a container on the control plane host(s).
\end{itemize}

Once the control plane has been initialized, the administrator may use the \emph{kubeadm} tooling to add additional workers to the cluster. However before the cluster can be utilized, it is essential that the administrator first pick and deploy an appropriate networking plugin~\cite{kubernetesnetworking}. In our community, the Calico\cite{calicodocs} plugin appears to be the most popular, given ease of setup/administration as well as a broad experience base.

\subsubsection{Declarative deployments with Kubespray}
The \emph{kubeadm} tool is sufficient for building and managing clusters, but often times a higher-level tool is useful for programmatic management of clusters. Building on the popular configuration management tool Ansible\cite{ansible}, the Kubespray tool  abstracts many of the deployment steps for a \emph{kubeadm}-based cluster on bare metal or virtual machines. With Kubespray, administrators can write a declarative YAML configuration to deploy a highly-available Kubernetes across a variety of Linux distributions with a sensible default configuration while retaining significant leeway in tools and plugin choices. Kubespray deployment is especially convenient in that it is designed for idempotence, such that repeated operations of the Kubespray ``playbook'' will result in the same cluster deployment, including resuming an interrupted deployment. Beyond installation, Kubespray offers convenient tooling for upgrading clusters between major versions of Kubernetes. 

One potential drawback of Kubespray is that the installation mechanism is push-based, relying on making outbound SSH connections to nodes in order to configure and manage them. The SSH-based push model may also be unsuitable for situations that require entire nodes to be dynamically added or removed from the cluster.

\subsection{OpenShift/OKD}
Red Hat's OKD is another downstream, fully compliant Kubernetes distribution that integrates a number of value-added components not found in the product published by the CNCF. Red Hat also sells a supported edition, \emph{OpenShift}, wherein they provide support for 4 minor releases of OpenShift, each having a 4 month release cycle, giving users at least 16 months of support for a given release \cite{openshift_lifecycle}. Red Hat has positioned OpenShift/OKD as an Enterprise solution for Kubernetes deployments with built-in features including: multi-tenancy, CI/CD, container registry, monitoring and log aggregation, metering, virtualization for traditional applications, and so on \cite{rh_openshift_kubernetes}. For the purposes of this article, we will refer to these two related distributions together as simply OKD. 

\subsection{CoreOS and Ignition}
Unlike other Kubernetes distributions, OKD can manage the entire lifecycle of a node in the cluster down to the baseboard management controller (via the Redfish API\cite{redfish}, an industry standard specification for RESTful management of servers). The OKD authors provide four ways to install OKD on bare-metal \cite{bare_metal_okd}, including an automated installer using Redfish, a web-based assistant built on top of the automated installer, an agent-based installer designed for restrictive or air-gapped networks, and a ``full control'' installation recipe that allows administrators to manage the cluster installation themselves. 

The OKD Kubernetes distribution is tightly integrated with Fedora CoreOS \cite{fedora_coreos}, a container-focused operating system that implements many of the Cloud Native Computing principles including immutability and declarative configuration at a low-level. Fedora CoreOS, like other CoreOS derivatives including Flatcar Linux\cite{flatcar_linux}, includes a relatively new server provisioning and initial configuration software \emph{Ignition}\cite{ignition}. Ignition uses a declarative configuration format to manage disks, filesystems, and services in CoreOS. OKD uses Ignition to create immutable server configurations, such that operating system updates are treated as an atomic process, with the server booting into an entirely new OS image loaded from the network after PXE.

By treating servers in a Kubernetes cluster as ephemeral objects much in the same way as Kubernetes treats containers in pods, the OKD distribution makes it easy to scale to integrate additional (or remove deprecated/repurposed) hardware as needed. However, because OKD assumes a high level of control over the server provisioning environment, it may be challenging in some instances to integrate it with existing site provisioning infrastructure. 


\subsection{Rancher}
Developed by SUSE, Ranchers occupies a space between the CNCF's upstream Kubernetes release and the highly-opinionated Openshift/OKD distribution. Rancher offers many features similar to OKD, including a web dashboard, options for deploying in especially security-conscious environments (such as air-gapped clusters), prescriptive choices of plugins including Helm and the Nginx Ingress Controller, etc. Like Red Hat, SUSE offers support contracts and licenses for Rancher that guarantee a certain level of Enterprise support for Kubernetes clusters. Rancher offers two versions of Kubernetes for different deployment targets: K3S for single-node clusters and edge networks, and RKE2 for a larger clusters and those that need strong security guarantees. 

\subsubsection{K3S}
Designed to be fully compliant with upstream Kubernetes, K3S\cite{k3s} is an opinionated Kubernetes distribution, built by the Rancher authors, that provides a base installation as well as a networking plugin (Flannel), DNS solution (CoreDNS), Ingress controller (Traefik), and other needed tools for a production instance. One unique feature of K3S is that it allows the administrator to use a traditional SQL database (e.g. SQLite, Postgres, MariaDB) instead of the embedded \emph{etcd} key-value database used by other Kubernetes distributions. However, K3S is explicitly not a fork of upstream Kubernetes and seeks to maintain compatibility with the core functionality while removing in-tree cloud and storage providers to shrink the final binary size. 

While it is possible to run larger sites on K3S, it seems especially suitable for sites with constrained available resources, as it is optimized for deployment on edge networks. One downside of K3S is that the default SQLite-based database is not particularly suitable for high-availability deployments. 

\subsubsection{RKE2 / Rancher Government}
The Rancher developers have also developed a Kubernetes distribution specifically targeted for compliance with the U.S. Government's ``Federal Information Processing Standard'' (FIPS): RKE2, also known as Rancher Government \cite{rancher_gov_rke2}. Compared to K3S, RKE2 has a stronger security posture and is DISA STIG-certified \cite{disa_rke2_stig} for environments with especially demanding security requirements. RKE2 more closely tracks the upstream CNCF Kubernetes releases than K3S, supports high availability with an odd-numbered quorum of control plane nodes, and uses the typical etcd database while more tightly bundling various disparate Kubernetes plugins and software.

\section{Comparison of features}
While we cover just three popular Kubernetes deployment methods observed in High Energy Physics computing, there are a significant number of other Kubernetes integrators working in the Cloud Native Computing space~\cite{nubenetes_matrix_table}, including the many downstream Cloud providers that offer proprietary solutions for popular Kubernetes features. Indeed, it is evident that there is no ``one size fits all'' solution for Kubernetes. To help sites decide on a distribution that is most appropriate for them, we will directly compare Kubespray using the official Kubernetes deployment tool kubeadm, OKD/OpenShift, and Rancher/K3S/RKE2 under a few different lenses. 

\subsection{Provisioning and deployment}

Table \ref{deployment} draws comparisons of the initial cluster deployment strategy. This is a key differentiator between Kubernetes distributions because different products have various scopes for managing the underlying operating system and server hardware. Note that OpenShift / OKD requires more nodes than other distributions for initial cluster bootstrapping. This is because a High Availability configuration is not considered optional under the OKD model, and it therefore does not support a single node control plane configuration. Sites should also be aware that Kubespray and OKD both require a node external to the cluster for the initial cluster provisioning. For Kubespray clusters, it would be wise to keep available a machine that can SSH to all nodes in the cluster for ongoing maintenance and operation. In the case of OKD, the documentation notes\cite{okd_bare_metal_ipi} that the provisioning machine can be removed after cluster initialization.
\begin{table}[hbtp]
\footnotesize
\centering
\begin{tabulary}{1.0\textwidth}{|L|L|L|L|}
\hline 
    \textbf{} & 
    \textbf{Kubespray} &
    \textbf{OpenShift/OKD} &
    \textbf{Rancher} \\
\hline
    \textit{Minimum number of nodes} &
    2 (SSH bootstrap + Kubernetes control plane) & 
    4 (Provisioner + 3 Kubernetes control planes) &
    1 (Kubernetes control plane)\\
\hline
    \textit{Deployment mechanism} &
    Ansible playbook via SSH &
    Web-assisted installer, commandline automated installer, agent-based installer for restricted networks, manual installation &
    Commandline utility\\
\hline
    \textit{High availability} &
    Optional &
    Required &
    Optional \\
\hline
    \textit{Linux distribution supported} &
    Several, c.f.\cite{kubespray} & 
    Fedora CoreOS & 
    Several, c.f.\cite{k3s}\\
\hline
    \textit{Networking plugin} &
    Several, c.f. \cite{cni} &
    OVN-Kubernetes (default), OpenShift SDN \cite{openshift-networking} & 
    Flannel (K3S), Several c.f. \cite{rke2-basic-network-options} (RKE2) \\
\hline
\end{tabulary}
\caption{Kubernetes Cluster Deployment mechanisms and requirements}
\label{deployment}
\end{table}

\subsection{Essential additional tools and operators}

After a cluster has been deployed, there are a number of additional plugins and integrations that a cluster administrator may need to deploy before the cluster is production-ready. The choice of which additional features to apply to a cluster is a key differentiator spanning the gamut of bare metal and cloud deployment options. 

\begin{itemize}
    \item \textbf{Persistent Storage}: If no persistent storage has been defined, then only ephemeral or stateless services can run. For certain types of caches or web services this may be appropriate, but in general some persistent storage is essential for databases, user data, etc. Sites typically use networked Rook \cite{rook_ceph} with Ceph storage or local storage devices for this purpose with various trade-offs. Red Hat's Openshift Data Foundation product is based on Rook and NooBaa (a multi-cloud object gateway), while Rancher's Longhorn\cite{longhorn} product delivers replicated block storage to services.
    \item \textbf{Load Balancer}: Services running in Kubernetes may have internally visible \emph{ClusterIPs} or may be presented externally on an ephemeral port via \emph{NodePorts}. For services which require a dedicated public IP on a standard port (e.g., web servers), a Load Balancer service is needed for this function. Some examples of general purpose Kubernetes load balancer software includes MetalLB\cite{metallbDocs} or PureLB\cite{pureLB}.
    \item \textbf{Application routing via Ingress}: The ingress controller operates at the application level to proxy internet-facing traffic (typically HTTP/HTTPS, but encapsulating and proxying generic TCP workloads is possible) to back-end services running in a cluster. Typically Nginx\cite{nginx} or Traefik\cite{traefik} are used as ingress controllers. OKD's ingress controller is based on HAProxy, while Rancher uses Nginx.  
    \item \textbf{Certificate Management}: Applications that are secured via TLS typically need a certificate issued that is recognized by standard Certificate Authorities. Typically tools like \emph{cert-manager}\cite{certManager} are used to automatically issue certificates through free authorities like Let's Encrypt\cite{letsEncrypt} using the ACME protocol \cite{acme_protocol}.
    \item \textbf{DNS Management}: Services running through Kubernetes will typically need DNS records created and assigned to internet-facing IPs in order to be usable in production. While it is possible to create these records manually, it is often cumbersome, especially if wildcard records are not appropriate for a given application. Instead, administrators will frequently use tools like ExternalDNS\cite{externalDNS} to programmatically manage the entire lifecycle of a DNS record including creation, update and deletion.
\end{itemize}

\subsection{Commonly used extras}

While not strictly necessary for production operations, many administrators add the following additional tools to their clusters:
\begin{itemize}
    \item \textbf{Safe secret storage}: In the default configuration \cite{kube_docs_secrets}, secrets in Kubernetes are somewhat misleadingly named as they are stored in plaintext at rest in the database and are merely base64 encoded for transport. Encryption at rest requires additional tooling \cite{kubernetes-encrypt-data} that is typically not enabled by default. Additionally, secrets that are stored in an external repository can be managed via tools like Sealed Secrets\cite{sealedSecrets} or SOPS\cite{sops}.
    \item \textbf{Monitoring and Alerting}: Kubernetes does not have any significant tooling for monitoring and alerting included in the base installation. Many administrators choose to use a tool like Prometheus\cite{prometheus} combined with Grafana\cite{grafana} to monitor and build alerts for their clusters. Tightly integrated distributions such as OKD and Rancher include Prometheus and Grafana by default.
    \item \textbf{GitOps}: Tools such as FluxCD\cite{fluxcd}, ArgoCD\cite{argocdDocs} allow Kubernetes objects to be managed via Git source code repositories for continuous delivery deployment patterns. While the base Kubernetes installation does not include a GitOps tool, OKD specifically offers OpenShift GitOps integration \cite{openshift-gitops} based on ArgoCD. Rancher meanwhile offers its own tool for GitOps, Fleet \cite{rancher-fleet}. 
    \item \textbf{Web Interface}: Minimalist installation methods like \emph{kubeadm} and Kubespray do not include a web interface by default, however the Kubernetes Dashboard\cite{kubernetes-web-ui-dashboard} can be added after installation. Tools like Rancher and OKD include built-in custom web interfaces. 
\end{itemize}

From Table \ref{integrations}, it is clear that many of the integrations commonly needed in production environments are not included with minimal deployments. However, this is not to say that they are not possible to install. Instead, these integrations simply do not come pre-installed to maximize flexibility. Some integrations, e.g. GitOps, are considered non-essential but come with optional add-on components such as Red Hat's OpenShift GitOps or SUSE's Fleet for OKD and Rancher, respectively. 

\begin{table}[ht]
\footnotesize
\centering
\begin{tabulary}{1.0\textwidth}{|L|L|L|L|}
\hline 
    \textbf{} & 
    \textbf{Kubespray} &
    \textbf{OpenShift/OKD} &
    \textbf{Rancher} \\
\hline
    \textit{Persistent Storage} &
    Not included & 
    OpenShift Data Foundation \cite{openshift-data-foundation} & 
    Longhorn \cite{longhorn} \\
\hline
    \textit{Load Balancer} &
    Not included & 
    MetalLB \cite{openshift-networking} & 
    Optional \cite{rancher-load-balancing} \\
\hline
    \textit{HTTPS Certificate Management} & 
    Not included &
    (Optional) cert-manager Operator \cite{openshift-cert-manager} &
    Not included \\
\hline
    \textit{Ingress Controller} &
    Not included &
    HAProxy-based \cite{okd-ingress-operator}&
    Nginx-based \cite{rancher-load-balancer-ingress} \\
\hline
    \textit{External DNS Management} &
    Not included &
    (Optional) External DNS Operator \cite{okd-external-dns-operator} &
    Not included \\
\hline
    \textit{Secret Encryption} &
    Not by default &
    Not by default \cite{okd-encrypting-etcd} &
    Yes, via AES-CBC \cite{rke2-secrets-encryption} \\
\hline 
    \textit{GitOps} &
    Not included &
    (Optional) OpenShift GitOps (ArgoCD) &
    Fleet \\
\hline
    \textit{Monitoring \& Alerting} &
    Not included & 
    Prometheus and Grafana (built-in) \cite{okd-monitoring-prometheus} &
    Prometheus and Grafana (built-in) \cite{rancher-monitoring-alerting}  \\
\hline
    \textit{Web Interface} & 
    Not included &
    Yes &
    Yes \\
\hline
\end{tabulary}
\caption{Additional Integrations for Production Environments}
\label{integrations}
\end{table}

\section{Conclusions}

The possible configuration space for Kubernetes deployments is immense, driven by its widespread popularity and adoption in many disciplines and enterprises, and its continuous evolution.  This diversity of options means there is no one-size-fits-all solution. Each deployment method offers advantages and trade-offs.  Site administrators will want to carefully weigh specific requirements and constraints against local requirements and restrictions. In our field, obviously security and resource sharing considerations may weigh heavily in the choice of deployment method and distribution.  We also encourage developers to take note of these differences, or more restrictive environments in some cases, when creating containerized applications and services for Kubernetes platforms.

\section{Acknowledgements}

This work was supported in part by several National Science Foundation awards, including OAC-2029176 \emph{Collaborative Research: IRNC: Testbed: FAB: FABRIC Across Borders}; OAC-1836650 \emph{Institute for Research and Innovation in Software for High Energy Physics (IRIS-HEP)}, and PHY-2120747, \emph{U.S. ATLAS Operations: Discovery and Measurement at the Energy Frontier}.

In preparing this note, we utilized ChatGPT~\cite{chatgpt} in an early draft, following the principles established in \emph{Nature 613, 612 (2023)}~\cite{nature_chatgpt}. Our methods included: 1) using the language model's re-write feature to enhance text clarity; 2) employing the reference look-up feature to identify optimal literature sources; 3) updating and verifying these sources for accuracy, and ensuring they aligned with our own judgments based on several years of practical deployment experience; and 4) engaging in contextual dialogs by posing specific questions to develop narratives about tool set adoption and breadth activities in the cloud-native landscape. In addition to adhering to the principles adopted by the editorial board of \emph{Nature}, we recognize transparency opinions noted in \cite{nature_d41586_023_00381_x} and ethical warnings highlighted in \cite{Dehouche2021}.

\newpage 
\setcounter{page}{1}
\bibliographystyle{unsrt}
\bibliography{bibliography}

\end{document}